\documentclass[12pt,a4paper]{article}
\usepackage{amsmath,amssymb}

\usepackage{amsmath}

\usepackage{color}
\usepackage[colorlinks,linkcolor=blue,citecolor=red]{hyperref}

\DeclareMathSymbol{\bbbr}{\mathalpha}{AMSb}{"52}
\DeclareMathSymbol{\bbbc}{\mathalpha}{AMSb}{"52}

\usepackage{mathtools}
\DeclarePairedDelimiter\abs{\lvert}{\rvert}

\begin{document}

\title{Kinetic equation for soliton gas: integrable reductions}

\author{E.V. Ferapontov$^{1, 2}$, M.V.Pavlov$^{3}$, }     
\date{}   
\maketitle     
\vspace{-5mm}
\begin{center}

$^1$Department of Mathematical Sciences \\ Loughborough University \\
Loughborough, Leicestershire LE11 3TU \\ United Kingdom \\
$^2$Institute of Mathematics, Ufa Federal Research Centre,\\
Russian Academy of Sciences, 112, Chernyshevsky Street, \\ Ufa 450077, Russia \vspace{5pt}\\
$^3$ Immanuel Kant Baltic Federal University\\
Institute of Physics, Mathematics and Informational Technology\\
Al. Nevsky St. 14, Kaliningrad, 236041, Russia\\
\ \\
e-mails: \\[1ex]
\texttt{E.V.Ferapontov@lboro.ac.uk}\\
\texttt{M.V.Pavlov@lboro.ac.uk}\\
\end{center}

\vspace{1cm}

\begin{abstract}

Macroscopic dynamics of soliton gases can be analytically described
by the thermodynamic limit of the Whitham equations, yielding an integro-differential kinetic equation for the
density of states.
Under a delta-functional  ansatz, the kinetic equation for soliton gas  reduces to a non-diagonalisable  system of hydrodynamic type whose matrix consists of several  $2\times 2$ Jordan blocks. Here we demonstrate the integrability of this system by showing that it possesses a hierarchy of commuting hydrodynamic flows and can be solved by an extension of the generalised hodograph method. Our approach is a generalisation of Tsarev's theory of  diagonalisable systems of hydrodynamic type  to quasilinear systems with  non-trivial Jordan block structure. 

\bigskip

\noindent MSC:  35Q51, 35Q83, 37K10.

\bigskip

\noindent
{\bf Keywords}: soliton gas, kinetic equation,  hydrodynamic reduction, commuting flow, conservation law, generalised hodograph formula. 
\end{abstract}

\newpage




\section{Introduction}

Quasilinear systems of the form
\begin{equation}\label{V}
u_t=V(u)u_x
\end{equation}
have been thoroughly investigated in the literature. Here $u=(u^1, \dots, u^n)^T$ is a column vector of the dependent variables and $V$ is a $n\times n$ matrix.  The main emphasis has always been on the strictly hyperbolic  case where the matrix $V$ has real distinct eigenvalues. Under the additional condition that the Haantjes tensor of matrix $V$ vanishes, any such system can be reduced to a diagonal  form,
\begin{equation}
r^i_t=v^i(r) r^i_x,
\label{r}
\end{equation}
$i=1, \dots, n$, in specially adapted coordinates $r^1, \dots, r^n$ known as Riemann invariants. Systems of  type (\ref{r}) govern a wide range of problems in pure and applied mathematics, see, e.g., \cite{Tsarev, Tsarev1, Dub, Serre, Sev}.  It was shown by Tsarev \cite{Tsarev, Tsarev1} that under the so-called semi-Hamiltonian constraint,
\begin{equation}\label{semi}
 \left(\frac{v^i_{r^j}}{v^j-v^i}\right)_{r^k}=\left(\frac{v^i_{r^k}}{v^k-v^i}\right)_{r^j},
\end{equation}
system (\ref{r}) possesses infinitely many commuting flows and conservation laws and can be solved by the generalised hodograph method (here $i\ne j\ne k$ and  $v^i_{r^j}=\partial v^i/\partial r^j$, etc). Recall that commuting flows  of diagonal systems (\ref{r}) are also diagonal,
\begin{equation*}
r^i_y=w^i(r) r^i_x,
\end{equation*}
where the requirement of commutativity,  $r^i_{ty}=r^i_{yt}$, implies the relations
$$
\frac{v^i_{r^j}}{v^j-v^i}=\frac{w^i_{r^j}}{w^j-w^i},
$$
$i\ne j$. Introducing the notation $a_{ij}=\frac{v^i_{r^j}}{v^j-v^i}$ and rewriting the equations for commuting flows in the form
\begin{equation}\label{co}
w^i_{r^j}=a_{ij}({w^j-w^i}),
\end{equation}
the requirement of their compatibility, $(w^i_{r^j})_{r^k}=(w^i_{r^k})_{r^j}$, implies the integrability conditions
\begin{equation}\label{semi1}
a_{ij, r^k}=a_{ij}a_{jk}+a_{ik}a_{kj}-a_{ij}a_{ik}
\end{equation} 
which are equivalent to the semi-Hamiltonian property (\ref{semi}). Under  conditions (\ref{semi1}), system (\ref{co}) for commuting flows possesses infinitely many solutions parametrised by $n$ arbitrary functions of one variable \cite{Tsarev}.

\bigskip

In this paper we study integrability aspects of quasilinear systems  (\ref{V}) whose matrix $V$ consists of $n$ Jordan blocks of size $2\times 2$:
\begin{eqnarray}
\label{J}
&&
\begin{array}{l} 
r^i_t=v^i r^i_x+p^i\eta^i_x,  \\
\eta^i_t=v^i\eta^i_x,  
\end{array}
\end{eqnarray}
$i=1, \dots, n$, where the coefficients $v^i(r, \eta)$ and $p^i(r,\eta)$ are functions of the $2n$ dependent variables $r=(r^1, \dots, r^n)$ and $\eta=(\eta^1, \dots, \eta^n)$. Systems of this form have appeared recently as delta-functional reductions of the kinetic equation for  soliton gas  \cite{PTE}, see below, as well as hydrodynamic reductions of linearly degenerate dispersionless integrable systems in multidimensions, see Section \ref{sec:Mik}. In Section \ref{sec:com} we  derive Jordan block analogues of equations for commuting flows (\ref{co}) and integrability conditions (\ref{semi1}). Although obtained in exactly the same way as for diagonalisable systems, these conditions are considerably more complicated.

\medskip

Block-diagonal systems of type (\ref{J}) share the following general properties:

\begin{itemize}

\item The Haantjes tensor of  system  (\ref{J}) vanishes identically, which makes  these systems natural parabolic analogues of diagonal  systems (\ref{r}). Note that systems (\ref{r}) can be obtained from (\ref{J}) by setting $\eta^i=const$.

\item The class of systems (\ref{J}) is invariant under changes of variables of the form
\begin{equation}\label{equiv}
r^i\to f^i(r^i, \eta^i), \qquad \eta^i\to g^i(\eta^i),
\end{equation}
where $f^i$ and $g^i$ are arbitrary functions of the indicated arguments.  Note that the group preserving the class of diagonal systems (\ref{r}) is more narrow, generated by transformations of the form $r^i\to f^i(r^i)$,  functions of one variable only.

\item Matrices $V$ corresponding to systems (\ref{J})  form a commutative family. In particular, hydrodynamic commuting flows of systems (\ref{J}) are also  of type (\ref{J}). This observation allows one to develop the integrability theory of such systems in full analogy with Tsarev's theory of diagonalisable systems (\ref{r}) \cite{Tsarev, Tsarev1}, by requiring the existence of a hierarchy of commuting flows.

\end{itemize}

\noindent {\bf Remark.} Integrable quasilinear systems of a single Jordan block type (of arbitrary size) have appeared in the literature as degenerations of hydrodynamic systems associated with multi-dimensional hypergeometric functions  \cite{KK}, in the context of parabolic regularisation of the Riemann equation \cite{KO2} and as   reductions of hydrodynamic chains and linearly degenerate dispersionless PDEs in 3D \cite{Pavlov1}. A connection of such systems  with the modified KP hierarchy was established in \cite{LXF}.

\bigskip

Our interest in systems (\ref{J}) stems from the study of El's integro-differential kinetic equation for  dense soliton gas \cite{El,EK, EKPZ}:
\begin{equation}\label{gas}
\begin{array}{c}
f_t+(sf)_x=0,\\
\ \\
s(\eta)=S(\eta)+\mathop{\int}_{0}^{\infty}G(\mu, \eta)f(\mu)[s(\mu)-s(\eta)]\ d\mu,
\end{array}
\end{equation}
where $f(\eta)=f(\eta, x, t)$ is the distribution function and $s(\eta)=s(\eta, x, t)$ is the associated transport velocity. Here the variable $\eta$ is the spectral parameter in the Lax pair associated with the dispersive hydrodynamics; the function $S(\eta)$ (free soliton velocity) and the kernel $G(\mu, \eta)$ (phase shift due to pairwise soliton collisions)  are independent of $x$ and $t$. 
The kernel $G(\mu, \eta)$ is assumed to be symmetric: $G(\mu, \eta)=G(\eta, \mu)$. Equation (\ref{gas}) describes the evolution of a dense soliton gas and represents a broad generalisation of Zakharov's kinetic equation for rarefied soliton gas \cite{Z}. It has appeared independently in the context of generalised hydrodynamics of multi-body quantum integrable systems \cite{Do}.
Note that both $S(\eta)$ and $G(\mu,\eta)$ are system specific. In the special case 
$$
S(\eta)=4\eta^2, \qquad G(\mu, \eta)=\frac{1}{\eta \mu} \log \abs[\Big]{ \frac{\eta-\mu}{\eta+\mu}},
$$
system (\ref{gas}) was derived in \cite{El} as thermodynamic limit of the KdV Whitham equations and generalised in \cite{EK, ET} to the NLS case. It was demonstrated in \cite{PTE} that under a delta-functional ansatz,
\begin{equation}\label{del}
 f(\eta, x, t)=\sum_{i=1}^{n}u^i(x, t)\ \delta(\eta-\eta^i(x, t)),
\end{equation}
system (\ref{gas}) reduces to a $2n\times 2n$ quasilinear system for $u^i(x, t)$ and $\eta^i(x, t)$,
\begin{equation}\label{uv}
u^i_t=(u^iv^i)_x, \qquad \eta^i_t=v^i\eta^i_x,
\end{equation} 
where $v^i$ can be recovered from the linear system
$$
v^i=-S(\eta^i)+\sum_{k\ne i}\epsilon^{ki}u^k(v^k-v^i), \qquad  \epsilon^{ki}={G(\eta^k, \eta^i)}, \ k\ne i.
$$
The special choice $\eta^i(x, t)=const$ was discussed previously in \cite{EKPZ}. In this case the last $n$ equations (\ref{uv}) are satisfied identically, while the first $n$ equations constitute an integrable diagonalisable linearly degenerate system. The main aim of this paper is to demonstrate integrability of the full system (\ref{uv}). This task  is not so straightforward since system (\ref{uv})  is not diagonalisable, thus,  Tsarev's theory of diagonalisable systems of hydrodynamic type (\cite{Tsarev, Tsarev1}, see also \cite{Dub}) does not apply, and a suitable generalisation  is  required. 
Following \cite{PTE}, let us introduce the new  variables $r^i$ by the formula
$$
r^i=-\frac{1}{u^i}\left(1+\sum_{k\ne i}\epsilon^{ki}u^k \right).
$$
In the dependent variables $r^i, \eta^i$, system (\ref{uv}) reduces to block-diagonal form (\ref{J})
 where the coefficients $v^i$ and $p^i$ can be expressed in terms of $(r, \eta)-$variables as follows. Let us introduce the $n\times n$ matrix $\hat \epsilon$ with diagonal entries $r^1, \dots, r^n$ (so that $\epsilon^{ii}=r^i$) and off-diagonal entries $\epsilon^{ik}={G(\eta^i, \eta^k)}, \ k\ne i$. 
 Note that this matrix is symmetric due to the symmetry of the kernel $G$. 
 Define another symmetric matrix $\hat \beta=-\hat \epsilon^{-1}$. Explicitly, for $n=2$ we have
 $$
 \hat \epsilon=\left(
 \begin{array}{cc}
 r^1&\epsilon^{12}\\
 \epsilon^{12}&r^2
 \end{array}
 \right), \qquad
  \hat \beta=\frac{1}{r^1r^2-(\epsilon^{12})^2}\left(
 \begin{array}{cc}
 -r^2&\epsilon^{12}\\
 \epsilon^{12}&-r^1
 \end{array}
 \right).
 $$
Denote $\beta_{ik}$ the matrix elements of $\hat \beta$ (indices $i$ and $k$ are allowed to coincide). Introducing the notation $\xi^k(\eta^k)=-S(\eta^k)$, we have  the following formulae for $u^i, v^i$ and $p^i$ \cite{PTE}:
\begin{equation}\label{vp}
u^i=\sum_{k=1}^n\beta_{ki}, \qquad v^i=\frac{1}{u^i}\sum_{k=1}^n \beta_{ki}\xi^k, \qquad
p^i=\frac{1}{u^i}\left(\sum_{k=1}^n \epsilon^{ki}_{,\eta^i}(v^k-v^i)u^k+(\xi^i)' \right),
\end{equation}
see Section \ref{sec:gas} for explicit form of $v^i$ and $p^i$ in the  simplest cases $n=1, 2$. Although the explicit form of $v^i$ and $p^i$ gets increasingly more complicated as $n$ grows, there exists a remarkably compact formula for the general solution  of system (\ref{J}) corresponding to reduction (\ref{uv}) that works for arbitrary $n$:
\begin{equation}\label{nn}
r^i=\frac{\varphi^i_{,\eta^i}-(\xi^i)'\, t}{\mu^i},\qquad
\varphi^i(\eta^1, \dots, \eta^n)=x+\xi^i(\eta^i)\, t;
\end{equation}
here $\mu^i(\eta^i)$ are arbitrary functions of their arguments and the functions $\varphi^i(\eta^1, \dots, \eta^n)$ satisfy the relations $\varphi^i_{,\eta^j}=\epsilon^{ji}(\eta^i, \eta^j)\, \mu^j(\eta^j)$, $i\ne j$, no summation. The last $n$ equations (\ref{nn}) define $\eta^i(x, t)$ as implicit functions of $x$ and $t$; then the first $n$ equations define $r^i(x, t)$ explicitly. 
Formula (\ref{nn}) results from the generalised hodograph approach outlined in Section \ref{sec:hod}.

We demonstrate integrability of system (\ref{uv}) by explicitly presenting its commuting flows (Section \ref{sec:com}) and conservation laws (Section \ref{sec:cons}) parametrised by  $2n$ arbitrary functions of one variable. 
We emphasise that integrability of system (\ref{uv})  holds for {\it arbitrary} functions $S(\eta)$ and $G(\mu, \eta)$ in the kinetic equation (\ref{gas}), even without  the assumption of symmetry of the kernel $G$. Finally, we point out that formula (\ref{nn}) does not  possess any obvious limit to the case $\eta^i=const$ discussed in \cite{EKPZ}.

\section{Commuting flows}
\label{sec:com}

Given system (\ref{J}), let us look for commuting flows in the form
\begin{eqnarray}
\label{cJ}
&&
\begin{array}{l} 
r^i_y=w^i r^i_x+q^i\eta^i_x,  \\
\eta^i_y=w^i\eta^i_x.  
\end{array}
\end{eqnarray}
The requirement of commutativity of (\ref{J}) and (\ref{cJ}), $r^i_{ty}=r^i_{yt}$ and $\eta^i_{ty}=\eta^i_{yt}$, leads to the two groups of  conditions. 
First of all, for every $i\in \{1, \dots, n\}$ one has the relations
\begin{equation}\label{c1}
\begin{array}{c}
\frac{v^i_{r^i}}{p^i}=a_i=\frac{w^i_{r^i}}{q^i},\\
\ \\
\frac{v^i_{\eta^i}-p^i_{r^i}}{p^i}=b_i=\frac{w^i_{\eta^i}-q^i_{r^i}}{q^i}.
\end{array}
\end{equation}
Secondly, for every  $i\ne j\in \{1, \dots, n\}$ one has the relations
\begin{equation}\label{c2}
\begin{array}{c}
\frac{v^i_{r^j}}{v^j-v^i}=a_{ij}=\frac{w^i_{r^j}}{w^j-w^i},\\
\ \\
\frac{v^i_{\eta^j}-a_{ij}p^j}{v^j-v^i}=b_{ij}=\frac{w^i_{\eta^j}-a_{ij}q^j}{w^j-w^i},\\
\ \\
\frac{p^i_{r^j}+a_{ij}p^i}{v^j-v^i}=c_{ij}=\frac{q^i_{r^j}+a_{ij}q^i}{w^j-w^i},
\\
\ \\
\frac{p^i_{\eta^j}+b_{ij}{p^i}-c_{ij}p^j}{v^j-v^i}=d_{ij}=\frac{q^i_{\eta^j}+b_{ij}{q^i}-c_{ij}q^j}{w^j-w^i}.
\end{array}
\end{equation}
Here $a_i, b_i, a_{ij}, b_{ij}, c_{ij}, d_{ij}$ in (\ref{c1}) and (\ref{c2}) is just the notation for the coefficients shared by  commuting flows of the hierarchy. Commutativity conditions (\ref{c1}), (\ref{c2}) are presented in symmetric form where the left-hand sides (respectively, right-hand sides) are expressed in terms of system (\ref{J}) (respectively, its commuting flow (\ref{cJ})).

\subsection{Integrability conditions}
\label{sec:int}

Integrability conditions of system (\ref{J}) can be obtained by calculating consistency conditions of the linear system (\ref{c1}), (\ref{c2}) governing commuting flows. Thus, we regard $v^i, p^i$ as given and $w^i, q^i$ as the unknowns.  First of all, for every $i\in \{1, \dots, n\}$ one has the relations
\begin{equation}\label{co1}
\begin{array}{c}
w^i_{r^i}=a_iq^i, \qquad w^i_{\eta^i}=q^i_{r^i}+b_iq^i.
\end{array}
\end{equation}
Secondly, for every  $i\ne j\in \{1, \dots, n\}$ one has the relations
\begin{equation}\label{co2}
\begin{array}{c}
w^i_{r^j}=a_{ij}({w^j-w^i}),\qquad
w^i_{\eta^j}=a_{ij}q^j+b_{ij}(w^j-w^i), \\
\ \\
q^i_{r^j}=-a_{ij}q^i+c_{ij}({w^j-w^i}), \qquad q^i_{\eta^j}=-b_{ij}{q^i}+c_{ij}q^j+d_{ij}({w^j-w^i}).
\end{array}
\end{equation}
Note that the relations $(w^i_{r^i})_{\eta^i}=(w^i_{\eta^i})_{r^i}$ give second-order derivatives $q^i_{r^i r^i}$:
\begin{equation}\label{qii}
q^i_{r^i r^i}=a_iq^i_{\eta^i}-b_iq^i_{r^i}+(a_{i, \eta^i}-b_{i, r^i})q^i.
\end{equation}
Expressed in terms of  the coefficients  $a_i, b_i, a_{ij}, b_{ij}, c_{ij}, d_{ij}$, the consistency conditions of relations (\ref{co1}), (\ref{co2}) and (\ref{qii})  are the required integrability conditions for system (\ref{J}). There are two types of integrability conditions, involving  two and three distinct indices, respectively. 

\medskip

\noindent{\bf Two-index conditions.} There are several groups thereof, each involving  two distinct  indices $i\ne j$. 

\noindent The consistency condition $(w^i_{r^i})_{r^j}=(w^i_{r^j})_{r^i}$ implies
\begin{equation}\label{I1}
a_{i, r^j}=0, \qquad a_{ij, r^i}=a_{ij}a_{ji}+a_ic_{ij}.
\end{equation}
The consistency condition $(w^i_{r^i})_{\eta^j}=(w^i_{\eta^j})_{r^i}$ implies
\begin{equation}\label{I2}
a_{i, \eta^j}=0, \qquad b_{ij, r^i}=b_{ij}a_{ji}+a_{ij}c_{ji}+a_id_{ij}.
\end{equation}
The consistency condition $(w^i_{\eta^i})_{r^j}=(w^i_{r^j})_{\eta^i}$ implies
\begin{equation}\label{I3}
b_{i, r^j}=2a_{ij}a_{ji}+2a_ic_{ij}, \qquad a_{ij, \eta^i}=a_{ij}b_{ji}-c_{ij}a_{ji}+b_ic_{ij}+c_{ij, r^i}.
\end{equation}
The consistency condition $(w^i_{\eta^i})_{\eta^j}=(w^i_{\eta^j})_{\eta^i}$ implies
\begin{equation}\label{I4}
b_{i, \eta^j}=2a_{ij}c_{ji}+2b_{ij}a_{ji}+2a_id_{ij}, \qquad b_{ij, \eta^i}=b_{ij}b_{ji}+a_{ij}d_{ji}-d_{ij}a_{ji}-c_{ij}c_{ji}+b_id_{ij}+d_{ij, r^i}.
\end{equation}
The consistency condition $(w^i_{r^j})_{\eta^j}=(w^i_{\eta^j})_{r^j}$ implies
\begin{equation}\label{I5}
a_{ij, r^j}=b_ja_{ij}-a_jb_{ij}-a_{ij}^2, \qquad a_{ij, \eta^j}=b_{ij, r^j}.
\end{equation}
The consistency condition $(q^i_{r^j})_{\eta^j}=(q^i_{\eta^j})_{r^j}$ implies
\begin{equation}\label{I6}
c_{ij, r^j}=b_jc_{ij}-a_jd_{ij}-2a_{ij}c_{ij}, \qquad c_{ij, \eta^j}=d_{ij, r^j}.
\end{equation}
Finally, direct calculation shows that the remaining consistency conditions between (\ref{qii}) and the last two equations (\ref{co2}), namely, $(q^i_{r^j})_{r^ir^i}=(q^i_{r^ir^i})_{r^j}$ and $(q^i_{\eta^j})_{r^ir^i}=(q^i_{r^ir^i})_{\eta^j}$, are satisfied identically. 
Relations (\ref{I1})--(\ref{I6}) form a complete set of integrability conditions for $4\times 4$ systems (\ref{J}) (case of two Jordan blocks). For more than two Jordan blocks, additional three-index conditions are required.

\medskip

\noindent{\bf Three-index conditions.} There are several groups thereof, each involving  three distinct  indices $i\ne j\ne k$. 

\noindent The consistency condition $(w^i_{r^j})_{r^k}=(w^i_{r^k})_{r^j}$ implies
\begin{equation}\label{I7}
a_{ij, r^k}=a_{ij}a_{jk}+a_{ik}a_{kj}-a_{ij}a_{ik}.
\end{equation} 

\noindent The consistency condition $(w^i_{r^j})_{\eta^k}=(w^i_{\eta^k})_{r^j}$ implies
\begin{equation}\label{I8}
\begin{array}{c}
a_{ij, \eta^k}=a_{ij}b_{jk}+a_{ik}c_{kj}+b_{ik}a_{kj}-a_{ij}b_{ik},
\ \\
b_{ij, r^k}=b_{ij}a_{jk}+a_{ik}b_{kj}+a_{ij}c_{jk}-a_{ik}b_{ij}.
\end{array}
\end{equation} 

\noindent The consistency condition $(w^i_{\eta^j})_{\eta^k}=(w^i_{\eta^k})_{\eta^j}$ implies
\begin{equation}\label{I9}
\begin{array}{c}
b_{ij, \eta^k}=a_{ij}d_{jk}+a_{ik}d_{kj}+b_{ij}b_{jk}+b_{ik}b_{kj}-b_{ij}b_{ik}.
\end{array}
\end{equation} 

\noindent The consistency condition $(q^i_{r^j})_{r^k}=(q^i_{r^k})_{r^j}$ implies
\begin{equation}\label{I10}
c_{ij, r^k}=c_{ij}a_{jk}+c_{ik}a_{kj}-c_{ij}a_{ik}-c_{ik}a_{ij}.
\end{equation} 

\noindent The consistency condition $(q^i_{r^j})_{\eta^k}=(q^i_{\eta^k})_{r^j}$ implies
\begin{equation}\label{I11}
\begin{array}{c}
c_{ij, \eta^k}=c_{ij}b_{jk}+c_{ik}c_{kj}+d_{ik}a_{kj}-a_{ij}d_{ik}-c_{ij}b_{ik},
\ \\
d_{ij, r^k}=d_{ij}a_{jk}+c_{ij}c_{jk}+c_{ik}b_{kj}-a_{ik}d_{ij}-c_{ik}b_{ij}.
\end{array}
\end{equation} 

\noindent The consistency condition $(q^i_{\eta^j})_{\eta^k}=(q^i_{\eta^k})_{\eta^j}$ implies
\begin{equation}\label{I12}
\begin{array}{c}
d_{ij, \eta^k}=c_{ij}d_{jk}+c_{ik}d_{kj}+d_{ij}b_{jk}+d_{ik}b_{kj}-b_{ij}d_{ik}-b_{ik}d_{ij}.
\end{array}
\end{equation} 
Relations (\ref{I1})--(\ref{I12}) form a complete set of integrability conditions for systems  (\ref{J}). They can be viewed as Jordan block analogues of the semi-Hamiltonian property (\ref{semi1}). For integrable systems (\ref{J}), the general solution of equations (\ref{co1}) -- (\ref{qii}) for commuting flows  depends on $2n$ arbitrary functions of one variable. 

For system (\ref{uv}) governing reductions of the soliton gas equation, all integrability conditions can be verified by direct calculation (which is quite cumbersome, see Section \ref{sec:gas}). In fact, in this particular case commuting flows can be found explicitly:   the general commuting flow (\ref{cJ}) of system (\ref{J}) corresponding to reduction (\ref{uv}) has the coefficients
\begin{equation}\label{cgas}
w^{i}=\frac{1}{u^{i}}\sum_{k=1}^n\beta _{ki}\varphi^k, \qquad q^{i}=\frac{1}{u^{i}}\left(\sum_{k=1}^n\epsilon^{ki}_{,\eta^i}(w^{k}-w^{i})u^k
-r^{i}\mu ^{i}+\varphi _{,\eta
^{i}}^{i}\right),
\end{equation}
where $\mu^i(\eta^i)$ are $n$ arbitrary functions of one variable and the functions $\varphi^i(\eta^1,\dots,\eta^n)$ satisfy the relations $\partial _{\eta ^{j}}\varphi ^{i}=\epsilon ^{ji}\mu^j,\text{ \ }j\neq i$, no summation (same functions as in  (\ref{nn})). The general commuting flow  depends on $2n$ arbitrary functions of one variable: $n$ functions $\mu^i(\eta^i)$, plus extra $n$  functions coming from  $\varphi^i$. This demonstrates  integrability of the system in question.

\section{Conservation Laws}
\label{sec:cons}

Conservation laws provide an alternative way to derive integrability conditions. Recall that conservation laws of  system (\ref{J}) are relations of the form
\begin{equation}\label{cons}
h_{t}=g_{x};
\end{equation}
here $h(r, \eta)$ is the conserved density and $g(r, \eta)$ is the corresponding flux. The requirement that  relation (\ref{cons}) holds
identically modulo (\ref{J}) implies the equations
\begin{equation*}
{ g}_{r^{i}}=v^{i}{ h}_{r^{i}}, \qquad {g}_{\eta ^{i}}=p^{i}{ h}_{
 r^{i}}+v^{i}{h}_{\eta ^{i}}.
\end{equation*}
Eliminating the flux $g$ we obtain a system of second-order linear PDEs for the density $h$:
\begin{equation}\label{int}
\begin{array}{c}
{h}_{r^{i}r^i}=b_{i}{ h}_{r^{i}}-a_{i}{ h}_{\eta ^{i}},\\
\ \\
{h}_{r^{i}r^j}=a_{ij}{ h}_{r^{i}}+a_{ji}{ h}_{r^{j}},\\
\ \\
{ h}_{r^{i}\eta^j}=a_{ji}{ h}_{\eta^j}+c_{ji}{ h}_{r^{j}}+b_{ij}{ h}_{r^{i}},\\
\ \\
{ h}_{\eta^i \eta^j}=d_{ij}{ h}_{r^{i}}+d_{ji}{ h}_{r^{j}}+b_{ij}
{ h}_{\eta^i}+b_{ji}{ h}_{\eta^j};
\end{array}
\end{equation}
here $i\ne j$ and the coefficients $a_i, b_i, a_{ij}, b_{ij}, c_{ij}, d_{ij}$ are the same as in Section \ref{sec:com}, which confirms the well-known fact that commuting flows share the same conserved densities. 
One can show that the compatibility conditions of  equations (\ref{int}) coincide with  integrability conditions (\ref{I1})--(\ref{I12}) obtained in Section \ref{sec:com}. For integrable systems (\ref{J}),  linear system (\ref{int})  is in involution and its general solution $h$ depends on $2n$ arbitrary functions of one variable. 

Note that system (\ref{uv}) possesses infinitely many conservation laws of the form
\begin{equation}\label{cn}
[u^if^i(\eta^i)]_t=[u^iv^if^i(\eta^i)]_x,
\end{equation}
where $f^i(\eta^i)$ are arbitrary functions of their arguments. Although these conservation laws involve only $n$ (rather than the required $2n$) arbitrary functions of one variable, their existence already implies the integrability of system (\ref{uv}), as well as of the corresponding system (\ref{J}).  This is a consequence of the following general proposition.

\medskip

\noindent {\bf Proposition 1.} Suppose that system (\ref{J}) possesses $2n$ conservation laws with functionally independent densities. Then the system is integrable (and automatically possesses infinitely many conservation laws parametrised by $2n$ arbitrary functions of one variable). 

\medskip

\centerline{\bf Proof:}

\medskip
This statement is a straightforward generalisation of the analogous fact known for diagonalisable systems: if system  (\ref{r}) possesses $n$ conservation laws with functionally independent densities, then it is semi-Hamiltonian (and, therefore, possesses infinitely many conservation laws parametrised by $n$ arbitrary functions of one variable), see \cite{Sev}. In the present context, the proof  can be summarised as follows. Note that all compatibility conditions of system (\ref{int}) for conserved densities $h$ have the form
$$
\sum_i\varphi_i h_{r^i}+\sum_i\psi_i h_{\eta^i}=0,
$$
where $\varphi_i$ and $\psi_i$ depend on the coefficients $a_i, b_i, a_{ij}, b_{ij}, c_{ij}, d_{ij}$ and partial derivatives thereof. We emphasize that the only `free' second-order partial derivative $h_{r^i\eta^i}$ will not enter any of these conditions.
Requiring that all  compatibility conditions are satisfied identically, that is, $\varphi_i=\psi_i=0$, we recover the full set of integrability conditions (\ref{I1})--(\ref{I12}). It remains to note that it is sufficient to require the existence of $2n$ functionally independent densities $h$ to reach the same conclusion, indeed, any non-trivial compatibility condition imposes a linear homogeneous relation on the gradient of $h$, thus contradicting functional independence. 

\medskip Setting in (\ref{cn}) $f^i=1$ and $f^i=\eta^i$, we obtain $2n$ functionally independent densities $u^i$ and $u^i\eta^i$. Thus, by the above proposition, system (\ref{uv}), as well as the corresponding system (\ref{J}), are integrable: all integrability conditions (\ref{I1})--(\ref{I12}) are satisfied identically. This  proof of integrability can be viewed as alternative to the one from  Section \ref{sec:int}.

In fact, all conservation laws of system (\ref{uv}) can be found explicitly:  the general conservation law  has the form
\begin{equation}\label{cn1}
\big[\sum_{i=1}^nu^i\psi^i(\eta)+\sum_{i=1}^n\nu^i(\eta^i)\big]_t=\big[\sum_{i=1}^nu^iv^i\psi^i(\eta)+\sum_{i=1}^n\tau^i(\eta^i)\big]_x;
\end{equation}
here $\nu^i(\eta^i)$ are arbitrary functions of one variable, the functions $\tau^i(\eta^i)$ can be recovered from the equations $(\tau^i)'=(\nu^i)'\xi^i$ and the functions $\psi^i(\eta^1, \dots, \eta^n)$ satisfy the equations
$\psi^i_{,\eta^j}=(\nu^j)'\epsilon^{ij}, \ j\ne i$. The general conservation law (\ref{cn1}) depends on $2n$ arbitrary functions of one variable: $n$ functions $\nu^i(\eta^i)$, plus extra $n$  functions coming from  $\psi^i$.
Setting in (\ref{cn1}) $\nu^i=0$ we recover conservation laws (\ref{cn}).

\section{Generalised hodograph formula}
\label{sec:hod}

Consider system (\ref{V}) together with its commuting flow,
$$
u_t=V(u)u_x \quad {\rm and} \quad u_y=W(u)u_x,
$$
where $V(u)$ and $W(u)$ are $n\times n$ matrices (the commutativity conditions $u_{ty}=u_{yt}$ impose differential constraints on $V$ and $W$). Then the matrix relation
\begin{equation}\label{imp}
W(u)=I\, x+V(u)\, t,
\end{equation}
where $I$ is the $n\times n$ identity matrix, defines an implicit solution $u(x, t)$ of system (\ref{V}) \cite{Tsarev, Tsarev1}. Note that, due to the commutativity conditions, only $n$ out of $n^2$ relations (\ref{imp}) will be functionally independent.
Since any integrable system (\ref{V}) possess infinitely many commuting flows parametrised by $n$ arbitrary functions of one variable, formula (\ref{imp}) provides a general (multivalued) solution.
For commuting systems (\ref{J}) and (\ref{cJ}), the hodograph formula (\ref{imp}) specialises to
\begin{equation}
\label{imp1}
w^i(r, \eta)=x+v^i(r, \eta)\, t, \qquad q^i(r, \eta)=p^i(r, \eta)\, t,
\end{equation}
which is a system of $2n$ implicit relations for the $2n$ dependent variables $r, \eta$. 

\medskip

\noindent{\bf Proposition 2.} Substituting into formula (\ref{imp1}) 
$v^i, p^i$ from (\ref{vp}) and $w^i, q^i$ from (\ref{cgas}), one obtains  formula (\ref{nn}) for the general solution of system (\ref{J}) corresponding to reduction (\ref{uv}) of the soliton gas equation:
$$
r^i=\frac{\varphi^i_{,\eta^i}-(\xi^i)'\, t}{\mu^i},\qquad
\varphi^i(\eta^1, \dots, \eta^n)=x+\xi^i(\eta^i)\, t.
$$

\medskip

\centerline{\bf Proof:}

\medskip

Using
$$
u^i=\sum_{k=1}^n\beta_{ki}, \qquad v^i=\frac{1}{u^i}\sum_{k=1}^n \beta_{ki}\xi^k, \qquad
w^{i}=\frac{1}{u^i}\sum_{k=1}^n\beta _{ki}\varphi^k,
$$
the relations $w^i=x+v^i\, t$ take the form
$$
\sum_{k=1}^n\beta_{ki}\varphi^k=x\sum_{k=1}^n \beta_{ki}+t \sum_{k=1}^n\beta _{ki}\xi^k,
$$
equivalently, 
$$
\sum_{k=1}^n\beta_{ki}(\varphi^k-x-\xi^k\, t)=0,
$$
which gives the second half of the required equations due to  non-degeneracy of matrix $\hat \beta$.

Similarly, using
$$
p^i=\frac{1}{u^i}\left(\sum_{k=1}^n \epsilon^{ki}_{,\eta^i}(v^k-v^i)u^k+(\xi^i)' \right), \qquad
q^{i}=\frac{1}{u^{i}}\left(\sum_{k=1}^n\epsilon^{ki}_{,\eta^i}(w^{k}-w^{i})u^k
-r^{i}\mu ^{i}+\varphi _{,\eta
^{i}}^{i}\right),
$$
the relations $q^i=p^i\, t$ take the form
$$
\sum_{k=1}^n\epsilon^{ki}_{,\eta^i}(w^{k}-w^{i})u^k-r^{i}\mu ^{i}+\varphi _{,\eta^{i}}^{i}=t\left(
\sum_{k=1}^n \epsilon^{ki}_{,\eta^i}(v^k-v^i)u^k+(\xi^i)'\right).
$$
Using $w^{i}=x+tv^{i}$, both sums in the above relation cancel out, leaving  
\begin{equation*}
-r^{i}\mu ^{i}+\varphi _{,\eta ^{i}}^{i}=t(\xi ^{i})^{\prime },
\end{equation*}
the first half of the required equations.

\section{Examples}
\label{sec:ex}

In Section \ref{sec:gas}  we give some more details on integrable system  (\ref{J}) coming from reductions of the kinetic equation for soliton gas. In Section \ref{sec:Mik} we discuss hydrodynamic reductions of type (\ref{J})  of the $3D$ Mikhalev system.

\subsection{Reductions of the kinetic equation for soliton gas}
\label{sec:gas}

Let us begin with the simplest cases $n=1$ and $n=2$.
\medskip

\noindent{\bf Case $n=1$.} In this case system (\ref{J}) is a single $2\times 2$ Jordan block,
\begin{eqnarray*}
&&
\begin{array}{l} 
r_t=\xi\, r_x-r\xi'\, \eta_x,  \\
\eta_t=\xi\, \eta_x,  
\end{array}
\end{eqnarray*}
recall that $\xi=-S(\eta)$. The general commuting flow of this system  has the form
\begin{eqnarray*}
&&
\begin{array}{l} 
r_y=\varphi\, r_x+(\mu r^2-\varphi'r)\, \eta_x,  \\
\eta_y=\varphi\, \eta_x,  
\end{array}
\end{eqnarray*}
where $\mu(\eta)$ and $\varphi(\eta)$ are two arbitrary functions. The corresponding generalised hodograph formula (\ref{imp1}) simplifies to
\begin{equation}\label{n=1}
r=\frac{\varphi'-\xi'\, t}{\mu}, \quad \varphi(\eta)=x+\xi(\eta)\, t;
\end{equation}
here the second relation specifies $\eta$ as an implicit function of $x$ and $t$,  the first relation gives an explicit formula for $r$.

\medskip

\noindent{\bf Case $n=2$.}  The corresponding  $4\times 4$ system (\ref{J}) has two Jordan blocks, setting $\epsilon^{12}=\epsilon$ in (\ref{vp}) we have
$$
v^1=\frac{r^2\xi^1-\epsilon \xi^2}{r^2-\epsilon}, \qquad v^2=\frac{r^1\xi^2-\epsilon \xi^1}{r^1-\epsilon},
$$
and
$$
p^1=\frac{\epsilon^2-r^1r^2}{r^2-\epsilon}\left(\frac{\xi^1-\xi^2}{r^2-\epsilon}\epsilon_{,\eta^1}+(\xi^1)'  \right), \qquad
p^2=\frac{\epsilon^2-r^1r^2}{r^1-\epsilon}\left(\frac{\xi^2-\xi^1}{r^1-\epsilon}\epsilon_{,\eta^2}+(\xi^2)'  \right),
$$
where $\xi^1=\xi^1(\eta^1), \ \xi^2=\xi^2(\eta^2)$ and $\epsilon(\eta^1, \eta^2)={G(\eta^1, \eta^2)}$. Direct computation of the invariants from Section \ref{sec:com} gives
$$
a_1=a_2=0, 
$$
$$
b_1=\frac{2r^2}{\epsilon^2-r^1r^2}, \qquad  b_2=\frac{2r^1}{\epsilon^2-r^1r^2},
$$
$$
a_{12}=\frac{r^1-\epsilon}{r^2-\epsilon}\ \frac{\epsilon} { r^1r^2-\epsilon^2}, \qquad a_{21}=\frac{r^2-\epsilon}{r^1-\epsilon}\ \frac{\epsilon} { r^1r^2-\epsilon^2},
$$
$$
b_{12}=\frac{\epsilon-r^2\frac{r^1-\epsilon}{r^2-\epsilon}}{r^1r^2-\epsilon^2}\ \epsilon_{,\eta^2}, \qquad
b_{21}=\frac{\epsilon-r^1\frac{r^2-\epsilon}{r^1-\epsilon}}{r^1r^2-\epsilon^2}\ \epsilon_{,\eta^1},
$$
$$
c_{12}=-\frac{r^1-\epsilon}{(r^2-\epsilon)^2}\ \epsilon_{,\eta^1}, \qquad
c_{21}=-\frac{r^2-\epsilon}{(r^1-\epsilon)^2}\ \epsilon_{,\eta^2},
$$
$$
d_{12}=\frac{r^1-\epsilon}{r^2-\epsilon}\ \epsilon_{,\eta^1 \eta^2}+\frac{r^1-r^2}{(r^2-\epsilon)^2}\ \epsilon_{,\eta^1}\epsilon_{ ,\eta^2}, \qquad
d_{21}=\frac{r^2-\epsilon}{r^1-\epsilon}\ \epsilon_{,\eta^1 \eta^2}+\frac{r^2-r^1}{(r^1-\epsilon)^2}\ \epsilon_{,\eta^1}\epsilon_{ ,\eta^2}; 
$$
note that the `soliton velocities' $\xi^1$ and $\xi^2$ do not enter these expressions. In fact, modulo equivalence transformations (\ref{equiv}), one can normalise $\xi^i(\eta^i)\equiv \eta^i$. Direct calculation shows that the corresponding $4\times 4$ system is integrable: it satisfies all two-index integrability conditions (\ref{I1})--(\ref{I6}). Note that the same conclusion is true in the more general case $G(\eta^1, \eta^2)\ne G(\eta^1, \eta^2)$, where the matrix $\epsilon$ is not symmetric.
\medskip

 For $n=2$, equations (\ref{co1}), (\ref{co2}) for commuting flows are  straightforward to solve. The general commuting flow is given by the formulae
\begin{equation}\label{wq2}
\begin{array}{c}
w^1=\frac{r^2\varphi^1-\epsilon \varphi^2}{r^2-\epsilon}, \quad w^2=\frac{r^1\varphi^2-\epsilon \varphi^1}{r^1-\epsilon},\\
\ \\
q^1=\frac{r^1r^2-\epsilon^2}{r^2-\epsilon}\left(\frac{\varphi^2-\varphi^1}{r^2-\epsilon}\epsilon_{,\eta^1}+r^1\mu^1- \varphi^1_{,\eta^1} \right), 
\quad
q^2=\frac{r^1r^2-\epsilon^2}{r^1-\epsilon}\left(\frac{\varphi^1-\varphi^2}{r^1-\epsilon}\epsilon_{,\eta^2}+r^2\mu^2- \varphi^2_{,\eta^2} \right);
\end{array}
\end{equation}
here $\varphi^1(\eta^1, \eta^2)$ and $\varphi^2(\eta^1, \eta^2)$ are two functions  such that
$\varphi^1_{,\eta^2}=\epsilon\, \mu^2$ and $\varphi^2_{,\eta^1}=\epsilon\, \mu^1$, where $\mu^1(\eta^1)$ and $\mu^2(\eta^2)$
are arbitrary functions of the indicated arguments. Thus, the general commuting flow depends on four arbitrary functions of a single argument: $\mu^1(\eta^1)$, $\mu^2(\eta^2)$, plus two more functions coming from the equations for $\varphi^1$ and $\varphi^2$. The corresponding generalised hodograph formula (\ref{imp1}) simplifies to
\begin{equation}\label{n=2}
\begin{array}{c}
\displaystyle{r^1=\frac{\varphi^1_{, \eta^1}-(\xi^1)'\, t}{\mu^1}, \quad r^2=\frac{\varphi^2_{, \eta^2}-(\xi^2)'\, t}{\mu^2}},\\
\ \\
\varphi^1(\eta^1, \eta^2)=x+\xi^1(\eta^1)\, t, \quad \varphi^2(\eta^1, \eta^2)=x+\xi^2(\eta^2)\, t;
\end{array}
\end{equation}
here the last two relations specify $\eta^1$ and $\eta^2$ as implicit functions of $x$ and $t$,  the first two relations give explicit formulae for $r^1$ and $r^2$. This is a special case of the general formula (\ref{nn}). 

For $n=2$,  equations (\ref{int}) for conserved densities are also straightforward to solve: a general  conserved density has the form
$$
h=\frac{(\epsilon-r^2)\psi^1+(\epsilon-r^1)\psi^2}{r^1r^2-\epsilon^2}+\nu^1(\eta^1)+\nu^2(\eta^2)
$$
where $\nu^1(\eta^1)$ and $\nu^2(\eta^2)$ are arbitrary functions of the indicated arguments and $\psi^1(\eta^1, \eta^2)$ and $\psi^2(\eta^1, \eta^2)$ are two functions  such that
$\psi^1_{,\eta^2}=(\nu^2)'\epsilon$ and $\psi^2_{,\eta^1}= (\nu^1)'\epsilon$. Thus, conserved densities depend on four arbitrary functions of one variable. Note that the above conserved density $h$ and the corresponding flux $g$ can be written in a simple form
\begin{equation}\label{hg2}
\begin{array}{c}
h=u^1\psi^1+u^2\psi^2+\nu^1(\eta^1)+\nu^2(\eta^2)\\
\ \\
g=u^1v^1\psi^1+u^2v^2\psi^2+\tau^1(\eta^1)+\tau^2(\eta^2),
\end{array}
\end{equation}
where $(\tau^1)'=(\nu^1)'\xi^1, \ (\tau^2)'=(\nu^2)'\xi^2$. Formulae (\ref{wq2}) and (\ref{hg2}) generalise to the case of arbitrary $n$ in a straightforward way, leading to the general commuting flows (\ref{cgas}) and the general conservation laws (\ref{cn1}); in fact, this is the way formulae (\ref{cgas}) and  (\ref{cn1}) were originally found.

\medskip

\noindent{\bf Case $n>2$.} Using computer algebra one can show that for $n=3$ and $n=4$, which correspond to $6\times 6$ and $8\times 8$ systems (\ref{J}), all integrability conditions (\ref{I1})--(\ref{I12}) are also satisfied identically (we do not present the explicit forms of $v^i$ and $p^i$ due to their complexity). 
 A general proof (alternative to the ones given in Sections \ref{sec:com} and \ref{sec:cons}) that works for arbitrary $n$ is based on the following explicit formulae for the coefficients
$a_i, b_i, a_{ij}, b_{ij}, c_{ij}, d_{ij}$:
$$
\begin{array}{c}
a_i=0, \quad b_i=2\beta_{ii}, \quad a_{ij}=\frac{u^{j}}{u^{i}}\beta _{ji}, \\
\ \\
 b_{ij}=\frac{\beta _{ji}}{u^{i}}\underset{p}{\sum }u^{p}\epsilon ^{pj}_{,\eta^j}+\frac{u^{j}}{u^{i}}\underset{p}{\sum }\beta_{pi}\epsilon ^{jp}_{,\eta^j}, \quad 
c_{ij}=\frac{u^{j}}{u^{i}}\underset{p}{\sum }\beta _{jp}\epsilon ^{pi}_{,\eta^i}-\frac{u^{j}}{(u^{i})^{2}}\beta _{ji}\underset{p}{\sum }u^{p}\epsilon ^{pi}_{, \eta^i},\\
\ \\
d_{ij}=\frac{u^{j}}{u^{i}}\epsilon
^{ji}_{,\eta^j\eta^i}+\frac{u^{j}}{u^{i}}{\underset{p}{\sum }}{
\underset{q}{\sum }}\epsilon ^{jq}_{,\eta^j}\beta
_{qp}\epsilon ^{pi}_{, \eta^i}+\frac{1}{u^{i}}{
\underset{p}{\sum }}\beta _{jp}\epsilon ^{pi}_{,\eta^i}{\underset{q}{\sum }}u^{q}\epsilon ^{qj}_{,\eta^j} \\
-\frac{u^{j}}{(u^{i})^{2}}{\underset{p}{\sum }}u^{p}\epsilon ^{pi}_{,\eta^i}{\underset{q}{\sum }}\epsilon ^{jq}_{,\eta^j}\beta _{qi}-\frac{\beta _{ji}}{(u^{i})^{2}}
{\underset{p}{\sum }}u^{p}\epsilon ^{pi}_{,\eta^i}
{\underset{q}{\sum }}u^{q}\epsilon ^{qj}_{,\eta^j}.
\end{array}
$$
Since all these coefficients are expressed in terms of $u^i$ and $\beta_{ij}$, the verification of integrability conditions (\ref{I1})--(\ref{I12}), which is a direct calculation,  requires the following differentiation rules:
\begin{equation*}
\beta _{ij, r^k}=\beta _{ik}\beta _{kj}, \qquad
\beta _{ij,\eta^k}=\beta _{kj}\underset{p}{\sum }\beta_{ip}\epsilon ^{pk}_{,\eta^k}+\beta _{ik}\underset{p}{\sum }\beta _{pj}\epsilon ^{kp}_{,\eta^k},
\end{equation*}
\begin{equation*}
u^{i}_{,r^j}=u^{j}\beta _{ji}, \qquad u^{i}_{,\eta ^{j}}=
\beta _{ji}{\underset{p}{\sum }}u^{p}\epsilon ^{pj}_{,\eta^j}+u^{j}{\underset{p}{\sum }
}\epsilon ^{jp}_{,\eta^j}\beta _{pi}.
\end{equation*}
Let us give a direct proof  that formula (\ref{nn}) provides a general solution of reduction (\ref{uv}) of  the soliton gas equation (which would not appeal to the generalised hodograph method).

\medskip

\noindent {\bf Proposition 3.} The general solution of system (\ref{J}) corresponding to reduction (\ref{uv}) of  the soliton gas equation is given by  formula (\ref{nn}):
$$
r^i=\frac{\varphi^i_{,\eta^i}-(\xi^i)'\, t}{\mu^i},\qquad
\varphi^i(\eta^1, \dots, \eta^n)=x+\xi^i(\eta^i)\, t;
$$
here $\mu^i(\eta^i)$ are arbitrary functions of their arguments and the functions $\varphi^i(\eta^1, \dots, \eta^n)$ satisfy the relations $\varphi^i_{,\eta^j}=\epsilon^{ji}(\eta^i, \eta^j)\, \mu^j(\eta^j)$, $i\ne j$, no summation.

\medskip

\centerline {\bf Proof:}

\medskip

Differentiating the last set of relations (\ref{nn}) by $x$ and $t$ we obtain
\begin{equation*}
\varphi _{,\eta ^{i}}^{i}\eta _{x}^{i}+\sum_{j\neq i}\mu ^{j}\epsilon
^{ji}\eta _{x}^{j}=1+t(\xi ^{i})^{\prime }\eta _{x}^{i}\quad \mathrm{and}
\quad \varphi _{,\eta ^{i}}^{i}\eta _{t}^{i}+\sum_{j\neq i}\mu ^{j}\epsilon
^{ji}\eta _{t}^{j}=\xi ^{i}+t(\xi ^{i})^{\prime }\eta _{t}^{i},
\end{equation*}
respectively. Using $\varphi _{,\eta ^{i}}^{i}-t(\xi ^{i})^{\prime
}=r^{i}\mu ^{i}$ we can rewrite these relations in the form
\begin{equation*}
r^{i}\mu ^{i}\eta _{x}^{i}+\sum_{j\neq i}\mu ^{j}\epsilon ^{ji}\eta
_{x}^{j}=1\quad \mathrm{and}\quad r^{i}\mu ^{i}\eta _{t}^{i}+\sum_{j\neq
i}\mu ^{j}\epsilon ^{ji}\eta _{t}^{j}=\xi ^{i}.
\end{equation*}
Taking into account that $\epsilon ^{ii}=r^{i}$, we come to
\begin{equation*}
\overset{n}{\underset{j=1}{\sum }}\mu ^{j}\eta _{x}^{j}\epsilon ^{ji}=1\quad 
\mathrm{and}\quad \overset{n}{\underset{j=1}{\sum }}\mu ^{j}\eta
_{t}^{j}\epsilon ^{ji}=\xi ^{i}.
\end{equation*}
Contracting these formulae (in the index $i$) with the inverse matrix  $\beta _{ik}$ and using $\epsilon^{ji}\beta_{ik}=-\delta^j_k$, we obtain
\begin{equation}\label{M}
-\mu^k\eta^k_x=\sum_{i=1}^n\beta_{ik}=u^k \quad {\rm and} \quad -\mu^k\eta^k_t=\sum_{i=1}^n\beta_{ik}\xi^i=u^kv^k,
\end{equation}
which readily implies $\eta _{t}^{k}=v^{k}\eta _{x}^{k}$.
\medskip

Similarly, differentiating $r^{i}\mu ^{i}=\varphi _{,\eta ^{i}}^{i}-t(\xi
^{i})^{\prime }$ by $x$ and $t$ we obtain (using $\varphi _{,\eta ^{i}\eta
^{j}}^{i}=\varphi _{,\eta ^{j}\eta ^{i}}^{i}=\epsilon _{,\eta ^{i}}^{ji}\mu
^{j}$):
\begin{equation*}
r_{x}^{i}\mu ^{i}+r^{i}(\mu ^{i})^{\prime }\eta _{x}^{i}=\varphi _{,\eta
^{i}\eta ^{i}}^{i}\eta _{x}^{i}+\sum_{j\neq i}\epsilon _{,\eta ^{i}}^{ji}\mu
^{j}\eta _{x}^{j}-t(\xi ^{i})^{\prime \prime }\eta _{x}^{i},
\end{equation*}
\begin{equation*}
r_{t}^{i}\mu ^{i}+r^{i}(\mu ^{i})^{\prime }\eta _{t}^{i}=\varphi _{,\eta
^{i}\eta ^{i}}^{i}\eta _{t}^{i}+\sum_{j\neq i}\epsilon _{,\eta ^{i}}^{ji}\mu
^{j}\eta _{t}^{j}-t(\xi ^{i})^{\prime \prime }\eta _{t}^{i}-(\xi
^{i})^{\prime },
\end{equation*}
respectively. Equivalently,
\begin{equation*}
r_{x}^{i}\mu ^{i}=\left(\varphi _{,\eta ^{i}\eta ^{i}}^{i}-r^{i}(\mu
^{i})^{\prime }-t(\xi ^{i})^{\prime \prime }\right)\eta _{x}^{i}+\sum_{j\neq
i}\epsilon _{,\eta ^{i}}^{ji}\mu ^{j}\eta _{x}^{j},
\end{equation*}
\begin{equation*}
r_{t}^{i}\mu ^{i}=\left(\varphi _{,\eta ^{i}\eta ^{i}}^{i}-r^{i}(\mu
^{i})^{\prime }-t(\xi ^{i})^{\prime \prime }\right)\eta _{t}^{i}+\sum_{j\neq
i}\epsilon _{,\eta ^{i}}^{ji}\mu ^{j}\eta _{t}^{j}-(\xi ^{i})^{\prime }.
\end{equation*}
Replacing $\eta _{t}^{i}$ by $v^{i}\eta _{x}^{i}$ and subtracting the first
of the above equations (multiplied by $v^{i}$) from the second, we obtain:
\begin{equation*}
(r_{t}^{i}-v^{i}r_{x}^{i})\mu ^{i}=\sum_{j\neq i}\epsilon _{,\eta
^{i}}^{ji}\mu ^{j}(v^{j}-v^{i})\eta _{x}^{j}-(\xi ^{i})^{\prime }.
\end{equation*}
To verify the equations $r_{t}^{i}=v^{i}r_{x}^{i}+p^{i}\eta _{x}^{i}$, it
remains to show that the following identity holds:
\begin{equation*}
p^{i}\eta _{x}^{i}\mu^i=\sum_{j\neq i}\epsilon _{,\eta ^{i}}^{ji}\mu
^{j}(v^{j}-v^{i})\eta _{x}^{j}-(\xi ^{i})^{\prime }.
\end{equation*}
This directly follows from the formula $\mu ^{i}\eta
_{x}^{i}=-u^{i}$, see (\ref{M}), indeed, under
this substitution the above formula reduces to
\begin{equation*}
p^{i}u^{i}=\sum_{j\neq i}\epsilon _{,\eta ^{i}}^{ji}(v^{j}-v^{i})u^{j}+(\xi
^{i})^{\prime },
\end{equation*}
which identically coincides with (\ref{vp}).

\medskip

\noindent {\bf Remark.} Reductions of the kinetic equation discussed in this paper satisfy the conditions $a_i=0$, which are equivalent to $v^i_{r^i}=0$.  These can be seen as Jordan block analogues of the conditions of {\it linear degeneracy} which are known to prevent breakdown of classical solutions. Thus, equation  (\ref{gas}) for soliton gas  should be viewed as a linearly degenerate kinetic equation, and is expected to have  good `regularity properties'.


\subsection{Reductions of the Mikhalev system}
\label{sec:Mik}

Integrable systems of type (\ref{J}), as well as integrable quasilinear systems (\ref{V}) with a more general Jordan block structure, typically arise as hydrodynamic reductions of multi-dimensional linearly degenerate integrable PDEs. As a simple illustration,
here we consider the linearly degenerate system 
\begin{equation}
\label{pav}
m_t=n_x, \qquad m_y= n_t-m n_x+n m_x, 
\end{equation}
which apparently first appeared in \cite{Mikhalev} in the context of Hamiltonian formalism of KdV type hierarchies and was subsequently investigated in the framework of multi-dimensional dispersionless integrability, see e.g.  \cite{Pavlov1, FM}.
Hydrodynamic  reductions  are sought in the form
\begin{equation}\label{mn}
m=m(r, \eta), \quad n=n(r, \eta),
\end{equation}
where the variables $r=(r^1, \dots, r^n)$ and $\eta=(\eta^1, \dots, \eta^n)$  satisfy a pair of commuting  systems  (\ref{J}), (\ref{cJ}).
Remarkably, all such reductions can be found explicitly: they are parametrised by $n$ arbitrary functions of one variable. 
The substitution of (\ref{mn}) into (\ref{pav}) gives
\begin{equation}\label{n}
n_{r^i}=v^im_{r^i}, \qquad n_{\eta^i}=v^im_{\eta^i}+p^im_{r^i},
\end{equation}
as well as the dispersion relations,
\begin{equation}\label{wq}
w^i=(v^i)^2-mv^i+n, \qquad q^i=(2v^i-m)p^i.
\end{equation}
Substituting (\ref{wq}) into commutativity conditions  (\ref{c1}) and using (\ref{n}) one obtains the relations
\begin{equation}\label{com2}
v^i_{r^i}=m_{r^i}.
\end{equation}
Substituting (\ref{wq}) into commutativity conditions  (\ref{c2}) and using (\ref{n}) one obtains the relations
\begin{equation}\label{com3}
v^i_{r^j}=m_{r^j}, \qquad v^i_{\eta^j}=m_{\eta^j}, \qquad p^i_{r^j}=0, \qquad p^i_{\eta^j}=0,
\end{equation}
$i\ne j$. Furthermore, calculating consistency conditions of equations (\ref{n}) we obtain
\begin{equation}\label{com4}
m_{r^i r^j}=0, \qquad m_{r^i \eta^j}=0, \qquad m_{\eta^i \eta^j}=0,
\end{equation}
$i\ne j$ (the last remaining consistency condition, $n_{r^i\eta^i}=n_{\eta^i r^i}$,  will be taken into account later). Equations (\ref{com2}), (\ref{com3}) and (\ref{com4}) are straightforward to solve: they imply 
$$
 m=\sum_kf^k(r^k, \eta^k), \qquad v^i=m+g^i(\eta^i),    \qquad p^i=p^i(r^i, \eta^i).
$$ 
Without any loss of generality, modulo equivalence transformations (\ref{equiv}), one can set $m=\sum_k r^k$, $v^i=m+\eta^i$. 
Then the last remaining consistency condition, $n_{r^i\eta^i}=n_{\eta^i r^i}$, implies $p^i_{r^i}=1$, which we will solve in the form
$p^i=r^i+({s^i}(\eta^i))'$. Ultimately, we arrive at the following formulae:
$$
m=\sum_k r^k, \qquad  n=\frac{1}{2}m^2+\sum_k r^k\eta^k+\sum_k s^k(\eta^k),  \qquad v^i=m+\eta^i, \qquad p^i=r^i+{(s^i(\eta^i))}';
$$
here $s^k(\eta^k)$ are arbitrary functions of their arguments. Integrability of these reductions can be verified by direct calculation.


\section{Concluding remarks}

The approach to hydrodynamic integrability  developed in this paper can be generalised to the whole class of non-semisimple quasilinear systems (\ref{V}) whose matrix $V$ has block-diagonal structure such that:
\begin{itemize}

\item each diagonal block has upper-triangular Toeplitz  form as in \cite{LXF}, with the unique eigenvalue; here is  a $3\times 3$ Toeplitz matrix with eigenvalue $v$:
$$
\left(\begin{array}{ccc}
v & p & q\\
0&v&p\\
0&0&v
\end{array}\right).
$$

\item different blocks have different eigenvalues. 

\end{itemize}

\noindent Matrices of this type (known as $\rm gl$-regular) form a commutative family, and the corresponding systems (\ref{V}) automatically  have zero Haantjes tensor. 
Commuting flows of such systems necessarily have  Toeplitz blocks of the same sizes.
An $n\times n$  system of this type should be called integrable if it possesses a hierarchy of hydrodynamic commuting flows/conservation laws parametrised by $n$ arbitrary functions of one variable.
In this connection, let us mention that a criterion for block-diagonalisability of quasilinear systems was obtained in  \cite{Bog} (theorems 1, 2).

\medskip Finally, it would be interesting to consider continuum versions of ansatz (\ref{del}),  system (\ref{uv}) and  its general solution (\ref{nn}), with an ultimate goal of establishing integrability of the full  kinetic equation (\ref{gas}).


\section*{Acknowledgements}

We thank A. Bolsinov, G. El and D. Tunitsky for useful discussions. The research of EVF was supported by a grant from the Russian Science Foundation No. 21-11-00006, https://rscf.ru/project/21-11-00006/. 
MVP was supported by the Ministry of Science and Higher Education of the Russian Federation (agreement no. 075-02-2021-1748).


\begin{thebibliography}{99}

\bibitem{Bog} O.I. Bogoyavlenskij, Block-diagonalizability problem for hydrodynamic type systems, J. Math. Phys. {\bf 47}, no. 6 (2006) 063502, 9 pp. 




\bibitem{Do} B. Doyon, T. Yoshimura,  Soliton gases and generalized hydrodynamics, J.-S. Caux, Phys. Rev. Lett. {\bf 120}
(2018) 045301.

\bibitem{Dub} B.A. Dubrovin, S.P. Novikov,
Hydrodynamics of weakly
deformed soliton lattices:
differential geometry and
Hamiltonian theory, Russian Math. Surveys,
{\bf 44}  (1989) 35-124.

\bibitem{EKPZ} G.A. El,  A.M. Kamchatnov, M.V. Pavlov, S.A. Zykov,  Kinetic equation for a soliton gas and its hydrodynamic reductions, J. Nonlinear Sci. {\bf 21}, no. 2 (2011) 151-191.

\bibitem{El} G.A. El,  The thermodynamic limit of the Whitham equations, Phys. Lett. A {\bf 311}, no. 4-5 (2003) 374-383.

\bibitem{EK} G.A. El,  A.M. Kamchatnov, Kinetic equation for a dense soliton gas, Phys. Rev. Lett., {\bf 95} (2005)  204101.

\bibitem{ET} G.A. El,  A. Tovbis, Spectral theory of soliton and breather gases for the focusing nonlinear Schr\"odinger equation, Phys. Rev. E {\bf 101} (2020) 052207. 

\bibitem{FP} E.V. Ferapontov,  M.V. Pavlov, Hydrodynamic reductions of the heavenly equation, 
Class. Quantum Grav. {\bf 20} (2003) 2429-2441.

\bibitem{FM} E.V. Ferapontov and J. Moss, Linearly degenerate PDEs and quadratic line complexes, Communications in Analysis and Geometry  {\bf 23}, no.1 (2015) 91-127.


\bibitem{KK} Yu. Kodama, B.G. Konopelchenko,  Confluence of hypergeometric functions and integrable hydrodynamic-type systems, Theoret. and Math. Phys. {\bf 188}, no. 3 (2016) 1334-1357.



\bibitem{KO2} B.G. Konopelchenko, G. Ortenzi,  Parabolic regularization of the gradient catastrophes for the Burgers-Hopf equation and Jordan chain, J. Phys. A {\bf 51}, no. 27 (2018) 275201, 26 pp.

\bibitem{LXF} Lingling Xue, E.V. Ferapontov, Quasilinear systems of Jordan block type and the mKP hierarchy,  J. Phys. A: Math. Theor. {\bf 53} (2020) 205202 (14pp).

\bibitem{Mikhalev} V.G. Mikhal\"ev, Hamiltonian formalism of Korteweg-de Vries-type hierarchies, Funct. Anal. Appl. {\bf 26}, no. 2 (1992) 140-142.

\bibitem{Pavlov1} M.V. Pavlov,  Integrability of exceptional hydrodynamic-type systems.  Proc. Steklov Inst. Math. {\bf 302}, no. 1 (2018) 325-335. 


\bibitem{PTE} M.V.  Pavlov, V.B. Taranov, G.A. El,  Generalized hydrodynamic reductions of the kinetic equation for a soliton gas, Theoret. and Math. Phys. {\bf 171}, no. 2 (2012) 675-682.





\bibitem{Serre} D. Serre, Systems of conservation laws. 1.
Hyperbolicity, entropies, shock waves, Cambridge University Press,
(1999) 263 pp;  Systems of conservation laws. 2.
Geometric structures, oscillations, and initial-boundary value problems,
Cambridge University Press (2000) 269 pp.

\bibitem{Sev} B. S\'evennec,  G\'eom\'etrie des syst\`emes
  hyperboliques de lois de conservation, M\'emoire (nouvelle s\'erie) N56,
  Suppl\'ement au Bulletin de la Soci\'et\'e Math\'ematique de France, {\bf
    122} (1994) 1-125.

\bibitem{Tsarev} {S.P. Tsarev,} Poisson brackets and one-dimensional
Hamiltonian systems of hydrodynamic type, Soviet Math. Dokl.
\textbf{31} (1985) 488-491. 

\bibitem{Tsarev1} {S.P. Tsarev,} The geometry of
Hamiltonian
systems of hydrodynamic type. The generalized hodograph method, Math. USSR
Izvestiya \textbf{37} (1991) 397-419.


\bibitem{Z} V.E. Zakharov, Kinetic equation for solitons, Sov. Phys. JETP {\bf 33}  (1971) 538-541.

\end{thebibliography}
\end{document}